\documentclass[twocolumn]{article}

\usepackage[english]{babel}

\usepackage[letterpaper,top=2cm,bottom=2cm,left=3cm,right=3cm,marginparwidth=1.75cm]{geometry}

\usepackage{amsmath}
\usepackage{graphicx}
\usepackage[colorlinks=true, allcolors=blue]{hyperref}
\usepackage{color}
\usepackage{authblk}
\usepackage{amsmath}
\usepackage[T1]{fontenc}
\usepackage[utf8]{inputenc}

\title{Muon radiography experiments on the subway overburden structure detection}
\author[1]{Xin Mao\thanks{*maoxin@buaa.edu.cn}}
\author[1]{Zhiwei Li \thanks{*zwli@whigg.ac.cn}}
\author[1]{Shuning Dong}
\author[1]{Jingtai Li}
\author[2]{Jianming Zhang}
\author[2]{Jie Pang}
\author[2]{Yaping Chen}
\author[3]{Bin Liu}
\author[3]{Xiaoping Ouyang}
\author[1]{Ran Han \thanks{*hanran@ncepu.edu.cn}}

\affil[1]{Beĳing Institute of Spacecraft Environment Engineering, 104 YouYi Road, Beĳing 100094, China}
\affil[2]{School of Nuclear Science \& Engineering, North China Electric Power University, Beĳing 102206, China}
\affil[3]{College of Nuclear Science and Technology, Beĳing Normal University, 19 Xinjiekou outer St., Beĳing 13 100875, China}

\begin{document}
\maketitle
\begin{abstract}
Muon radiography is an innovative and non-destructive technique for internal density structure imaging, based on measuring the attenuation of cosmic-ray muons after they penetrate the target. Due to the strong penetration ability of muons, the detection range of muon radiography can reach the order of hundreds of meters or even kilometers. Using a portable muon detector composed of plastic scintillators and silicon photomultipliers, we performed a short-duration(1h) flux scanning experiment of the overburden above the platform and tunnel of the Xiaoying West Road subway station under construction. With the observation direction facing up, the detector is placed on the north side of the track and moved eastward from the platform section inside the station to the tunnel section. The scanning length is 264m and a total of 21 locations are observed. By comparing the observed and predicted values of the muon 
survival ratio at different locations, the experiment accurately detects the anomalous density loss caused by the lobby on the first floor of the platform section. Furthermore, unknown anomalies caused by random placed light brick piles and side passage entrance above the observation locations are detected and confirmed later. This experiment verifies the feasibility of using natural muons to quickly detect abnormal structures of the upper layer of tunnel, and shows that muon radiography has broad application prospects in tunnel safety and other similar aspects.
\end{abstract}

\section{Introduction}
By the end of 2021, China's total subway mileage has ranked first in the world with 7253.73 $km$, and the construction of municipal subway tunnels is also in a period of rapid development\cite{china1}, therefore, the safety of tunnel construction and operation cannot be ignored. The compactness and stability of the rock and soil above the tunnel are directly related to the safety of the subway construction and operation. Quicksand, cavities or water bladders formed in the overburden due to reasons such as leakage of urban underground pipelines can put subway tunnels at risk of water inrush and collapse\cite{pipeleakage}. Between 2003 and 2016, there were 49 subway construction collapse accidents in China, including 28 collapse accidents caused caused by pipeline leakage\cite{Subwayrisk}. Reasonable internal inspection of the upper layer of the tunnel and timely detection of the abnormal density structure can help avoid risks in the construction and operation of the subway in advance.

Traditional geophysical methods, including electromagnetism, seismic and gravity have their own advantages and limitations. The electromagnetic method can realize the detection of geological bodies by obtaining the conductivity structure of the underground medium, which has the advantages of large detection depth and high efficiency, but is insensitive to the density structure and is easily affected by large metal structures or mineralized zones\cite{electro}. The seismic method has a large detection range and can obtain high-resolution wave velocity structures, but the reliability of the density structure obtained according to the wave velocity-density empirical relationship is affected to a certain extent, and it needs to rely on natural or artificial seismic sources, which will have a certain impact on the target structure\cite{Shapiro2005High,2017Crosshole}. The gravitational method has problems such as low spatiotemporal resolution, insufficient vertical resolution, and strong multi-solution\cite{2019gravity}. In contrast, Muon radiography, also called muography in brief, has the characteristics of high precision, large detection depth, non-destructive by using of natural sources, and can obtain the internal density structure of the detection target, which is more suitable for urban subway tunnels, ancient buildings and other large targets that are not suitable for intrusion and destruction.

Muons are secondary particles produced by the interaction of high-energy cosmic rays from the universe, mainly protons, with atomic nuclei in the atmosphere, with an average lifetime of 2 microseconds and traveling at the speed of light. The muons reaching sea level have an energy of more than 1 $GeV/c$ and a flux of about $170Hz/m^2$\cite{2016Review}. On the contrary, the number of other particles such as protons, mesons and electrons reaching the surface can be ignored. Muons will undergo a process of energy loss and deflection of the incident direction during the penetration of the target, due to inelastic collisions with electrons and elastic scattering from nuclei\cite{1977COSMIC}. High-energy muons can penetrate hundreds or even thousands of meters of rock formations\cite{O1985Cosmic} while low-energy muons will be stopped. The density structure inside the target can be indirectly obtained by measuring the attenuation of the muon flux before and after passing through the target object, which is the basic principle of muography.

Research on muography began in 1950 and was first applied in 1955 to measure overburden thickness above underground laboratories\cite{1955Cosmic}. With the improvement of detector level and electronic technology, muography has been successfully applied to geology and archaeology\cite{2008Radiographic,2022Imaging,Han_2020,2019Bedrock,2017Discovery}. Muography is more suitable for detecting internal chambers in large buildings or underground structures. In 2017, an unknown cave is detected by performing a muography experiment 40 meters underground in Mt. Echia, Naples\cite{2017Imaging}. In the same year, an unknown space hidden inside the Pyramid of Khufu is discovered by the same technique\cite{2017Discovery}. In 2020, a muon flux observation experiment of the 770m-long Alfreton tunnel in the United Kingdom was conducted and an unknown overburden void was identified, which was subsequently officially confirmed\cite{PhysRevResearch3017}.

In order to verify the feasibility of muography in detecting structural anomalies of the overburden of the tunnel and explore the potential of the muography in engineering applications, we conducted a muon observation experiment of the overburden of the tunnel at the Xiaoying West Road subway station under construction. We analyzed the abnormalities in the overburden layer, and derived the average density distribution and average thickness distribution. In this paper, after an introduction on the principles of determining anomalies, the detection system and experimental process will be presented, and finally the experimental results will be analyzed and summarized.

\section{Methods}
\subsection{Determination method of anomalies}
The degree of attenuation of muons through the substance is defined as the ratio of the flux reaching the detector after penetrating the overburden (underground case) to the amount of flux reaching the detector before penetrating the overburden (open-sky case), here referred to as survival/transmission ratio, expressed by ratio, abbreviated as R. The principle for judging the existence of anomalies is that the observation value of ratio is deviated from the prediction value of ratio without abnormal structure, also the deviation is outside the statistical error range of the data.

When there is no anomalous structure inside the object, the number $N$ of muons from a specific direction $\theta_{0} $ detected by the detector in $\Delta T $ time can be calculated by the following formula:
\begin{equation}
N(\theta_{0})=\tau(\theta_{0})\cdot\Delta{T}\cdot\varepsilon\cdot\int^{\infty}_{E_{min}(\rho)}\phi(\theta_{0},E)dE \label{e1}
\end{equation}
Among them, $\tau(\theta_{0})$ is the geometric acceptance of the muon detector, representing the influence of the detector geometry on the detection performance, the unit is $cm^{2}\cdot{sr}$, depending on the number of detector pixels $N_x\cdot{N_y}$, pixel size $d$, and the separation distance $D$ of the top and bottom layers of the detector.  $\tau(\theta_{0})$ can be obtained by integrating the detection area with solid angular resolution\cite{2022Imaging,Han_2020}. The geometric acceptance in different directions is different and is the largest in the vertical direction. Futhermore, the greater the deviation from the vertical direction, the smaller the acceptance. $\Delta{T}$ is the detection time, in second. The $\varepsilon$ is the detection efficiency, which can be obtained by scaling the efficiency of each scintillator unit\cite{Han_2020}.  $\phi(\theta_{0},E)$ is the differential energy spectrum of the incident muon in $cm^{-2}\cdot{sr^{-1}}\cdot{s^{-1}}\cdot{GeV^{-1}}$, and when predicting the muon flux, it is usually approximated by using empirical formulas obtained by fitting the muon data measured at sea level, such as Gaisser's formula [27][28], or directly using some software packages that simulate the interaction of primary cosmic rays with the atmosphere, such as Sybill\cite{PhysRevD.80.094003}, CRY\cite{Hagmann2007CosmicraySG}, FLUKA\cite{BATTISTONI2007286}, etc. $E_{min}$ is the minimum energy required for the muon to penetrate a substance of a certain thickness, depending on the density length $\rho$ of the penetrating object. $\rho$ is defined as the integral of the density of matter penetrated by the muon along the path, or can be approximated by multiplying the average density along the path by the length of the path. The relationship between the $E_{min}$ and $\rho$ in the energy range of $10MeV\sim 100TeV$ in different common substances can be obtained by looking up the summarized tables\cite{2001ADNDT..78..183G}, or by simulating with Monte Carlo simulation tools such as Geant4.

In the absence of an anomalous structure inside the object, within the $\theta$ cone angle, the predicted value of the muon survival ratio can be expressed by the following formula.
\begin{equation}
R(\theta)=\frac{\int_{\theta}N(\theta_{0})d\theta_{0}}{\int_{\theta}N_{opensky}(\theta_{0})d\theta_{0}}=\frac{\int_{E_{min}(\rho)}\phi(\theta_{0},E)dE}{\int_{E_{0}}\phi(\theta_{0},E)dE}\label{e2}
\end{equation}
Among them, $E_{0}$ is the minimum energy threshold of the muon that the detector can detect in the open sky case. It can be seen from Equation \eqref{e2} that using the survival ratio to judge the anomaly of the density structure can eliminate the influence of the detector effect on imaging.

\subsection{The relationship between the ratio and density length}
\begin{figure}[htbp]
\centering
\includegraphics[width=0.5\textwidth]{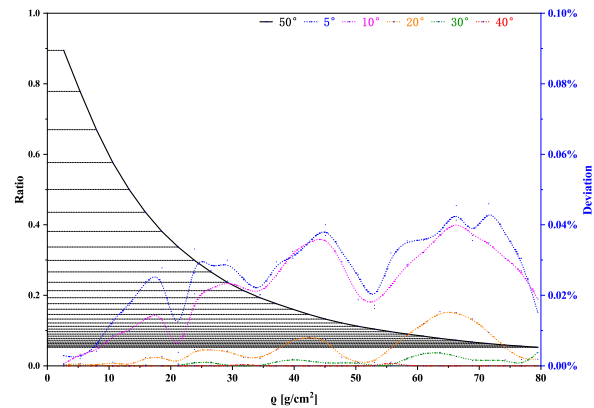}
\caption{\label{fig:ratio-rho}The relationship between ratio and density length. (solid black line: ratio of muons within $50^\circ$ cone angle; Blue, pink, yellow, green, red dotted line: $5^\circ$, $10^\circ$, $20^\circ$, $30^\circ$, $40^\circ$ relative deviation from the case of $50^\circ, {\Delta(\theta)=\frac{|R(\theta)-R(50^{\circ})|}{R(50^{\circ})}}$.}
\end{figure}

According to the definition of density length, if its relationship to the ratio is known, the average value of either quantity in density and length can be derived by assuming in advance that the other quantity is constant. To get the relationship, the CRY software package is used to generate cosmic ray muons, and Geant4 is used to simulate the interaction between muons and matter, and finally the relationship between ratio and penetration density length is obtained. CRY is short for Cosmic-ray shower generator, and the CRY software package can generate the distribution of cosmic ray particle beams at three altitudes, that is sea level, 2100m and 11300m, as a simulated input to the detector response. Geant4 is a Monte Carlo simulation tool commonly used to simulate the transportation and interaction of particles in matter, and is widely used in high-energy physics, nuclear physics, medicine, aerospace and other fields. The main process of this method is to establish the CAD model of the target, import it into Geant4, use the muon source generated by CRY, establish the detector in Geant4, select the built-in physical process in Geant4 for simulation, and finally obtain the muon flux rate received by the detector. Based on this method, the penetration of muons through different thicknesses of concrete of which the average density is assumed to be $2.3g/cm^3$ is simulated, and the ratios of muons received in different angle ranges are obtained as shown in Figure \ref{fig:ratio-rho}. The solid black line in the figure represents the relationship between the ratio of the muons in the range of 50° cone angle and the density length of the penetrated object, and the remaining dashed lines represent the relative deviation of the ratio value of muons in the range of other cone angles to the result of 50°, which is represented by the following formula.
\begin{equation}
    \Delta(\theta)=\frac{|R(\theta)-R(50^{\circ})|}{R(50^{\circ})} \label{e3}
\end{equation}

\section{Experiment and results}
\subsection{Detection system}

\begin{figure}[htbp]
\centering
\includegraphics[width=0.5\textwidth]{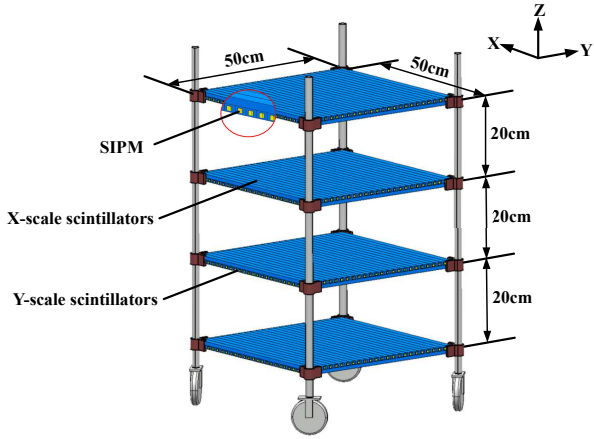}
\caption{\label{fig:detector}Schematic diagram of the structure of the detector.}
\end{figure}

The detector, based on plastic scintillators and silicon photomultiplier devices (SIPM), is an upgraded version of the previous generation [20][19][21]. The main improvements are higher position resolution, higher detection efficiency, smaller size and less power consumption, and more portable and convenient for moving and observing in complex scenes. As shown in Figure \ref{fig:detector}, the detector consists of four layers of detection unit plates arranged equidistant vertically to form $Z$ coordinates, and the plate spacing is $20cm$. Each layer consists of two $2500cm^2$ planes, each consisting of 25 plastic scintillator arrangements, each coupled with a single-ended SIPM. The scintillators of the two planes are orthogonally arranged together to form $X$ and $Y$ coordinates. The geometric size of the detector determines that its angular aperture of about $79^{\circ}$, and the maximum acceptance is $2.78cm^{2}sr$. The bottom of the detector is mounted with pulleys for convenient movement. In addition, the detector consumes $20W$ and is equipped with a $2000Wh$ mobile lithium battery, thus can continuously operate in places without mains power.

The most basic unit of the detector is the strip plastic scintillator with a rectangular cross-section. When the muon hits the scintillator, there will be ionization, the atomic deexcitation will produce a light signal, and the light signal will be detected by SIPM when it propagates along the scintillator to the side. The scintillator density used is $1.02g/cm^3$, the light output relative to anthracene crystals is $50\%\sim60\%$, the emission wavelength is between $395 nm\sim425 nm$, the attenuation time is $2.4 ns$, the attenuation length is $2.1m$, and the measurement range is between $50 keV\sim{3 MeV}$. Each scintillator is $2 cm\times1 cm \times 50 cm$ in size, resulting a high average angular resolution of $33.3mrad$, and is coated with a silver reflective film (ESR), which can increase the reflectivity of photons inside and shield foreign light sources to improve the efficiency of photon collection. At the scintillator end face, the J series SIPM sensor produced by onsemi is used to detect the optical signal, the size is $6 mm\times 6mm$, and it has a high quantum efficiency (PDE) for weak light, the maximum is $50\%$, and the corresponding wavelength is $420 nm$, which corresponds to the maximum emission wavelength of the scintillator.

\begin{figure}[htbp]
\centering
\includegraphics[width=0.5\textwidth]{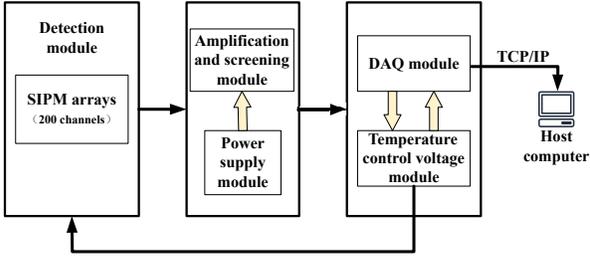}
\caption{\label{fig:detectionscheme}Block diagram of the detection system.}
\end{figure}

The front-end and back-end electronics (FEE) is used to process and upload the electrical signal, converted by SIPM from the detected optical signal, in 16-base form via TCP/IP protocol. Finally the data is analyzed and stored by the host computer software. The FEE consists of 8 amplification screening modules and 1 data acquisition board. Each amplification screening module can perform year-over-year amplification and over-threshold screening of 25 signals, and if the threshold of $32mV$ is exceeded after triggering signal amplification, it is considered "fire", and the over-threshold screening circuit then outputs a $3.3 V$ TTL level signal to the detector data acquisition board. The acquisition board is based on the Xilinx ZYNQ7035 series FPGA (Field Programable Gate Array), which can simultaneously read a total of 200 level signals from four layers of detection units. In addition, the board integrates multi-function peripherals such as temperature control voltage module, FPGA start-stop control module, and data transmission interface module, which can provide SiPM bias and temperature compensation for the detector at the same time. The sampling frequency of the acquisition card is $100MHZ$. When the optical signal is transmitted from the hitting point position to the SIPM detection, there is a delay in time. In order to prevent the signal acquisition is incomplete, the signal collected in 20 clock cycles is taken by the same logical bit "or" into a "real signal". When the real signals of X and Y "fire" at the same time, the acquisition board will package the detection unit number, temperature, signal into a 16-base data of 19 byte and upload it to the host computer through the TCP/IP protocol. Finally, the host computer selects, displays and saves the data of the muon events. Among them, the case can be considered as muon case only if the hitting time of the four detection units are in the same window ($200ns$) and the hitting points of the four detection units are in the same straight line. The detection framework of the system is shown Figure \ref{fig:detectionscheme}.

\begin{figure*}[ht]
\centering
\includegraphics[width=0.8\textwidth]{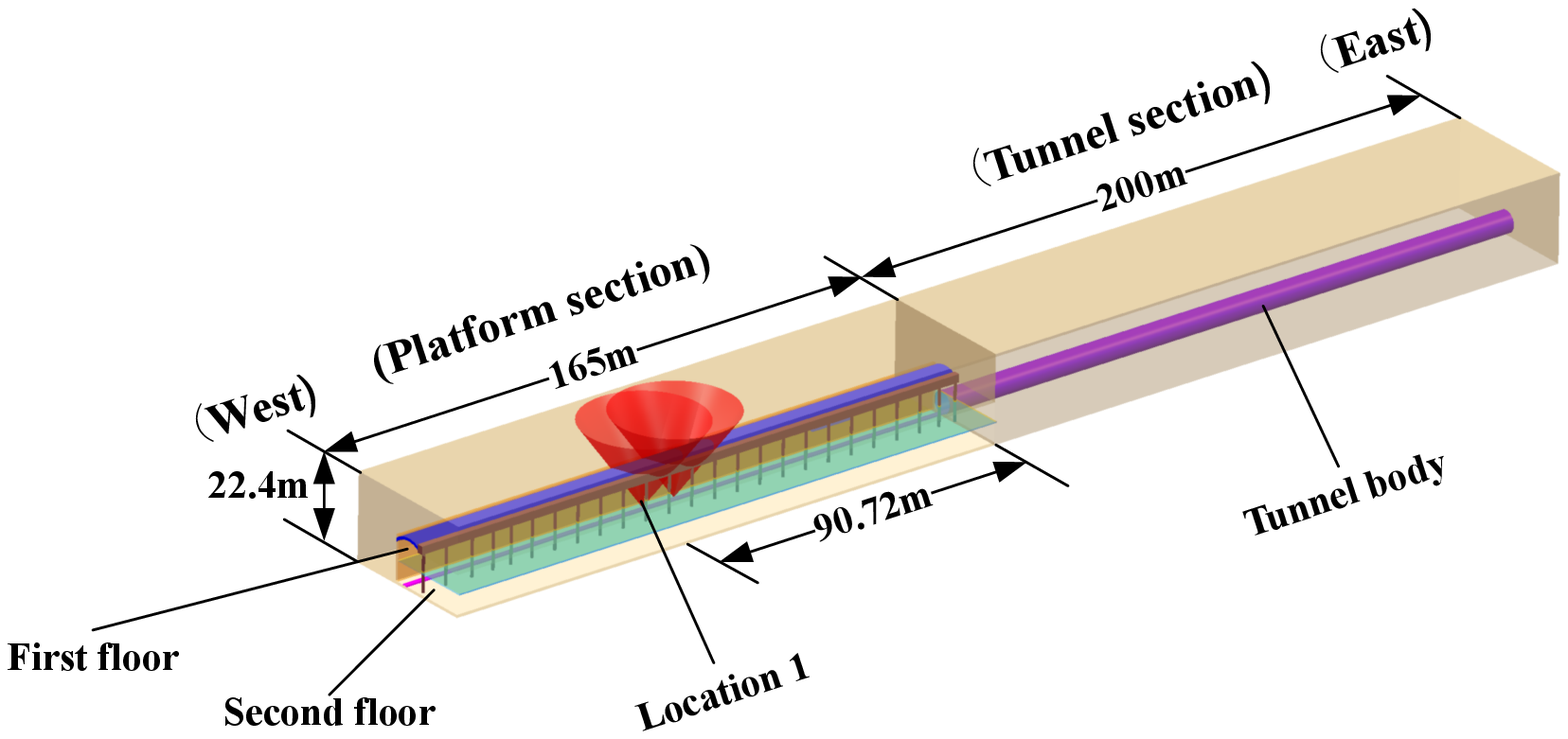}
\caption{\label{fig:CAD}1:1 CAD model of the Xiaoying West Road subway station and tunnel for Geant4 simulation.}
\end{figure*}

\subsection{Experiment}
The experimental site is located at Xiaoying West Road subway station in Beijing. The station has two reverse lines, the left line is on the north side and heads from west to east, the right line is on the south side and heads from east to west. This experiment is located on the left line. The subway station and tunnel of the left line are drawn at the actual 1:1 scale CAD model shown in Figure \ref{fig:CAD}. The subway track extends from west to east through the subway station to the tunnel and is represented in pink. The platform inside the station is $165 m$ long (platform section), and the tunnel outside the station is about $1.17 km$ (tunnel section) east to the next station. Inside the station is a two-story hall, the top of the hall is an arch with a radius of $7.6m$ (dark blue in the picture), the vault is $0.8m$ thick, the thickness of the overburden between the vault and the ground is $8.7m$, between the two floors of the hall is a $0.4m$ thick concrete floor, and the hall is arranged with 26 marble columns with a diameter of $0.8m$ along the east-west platform. On the north edge of the hall is a $0.8m$ thick concrete wall (brown in the picture). Outside of the hall, the shield of the tunnel part has an inner diameter of $2.7m$ and a lining thickness of $0.3m$, which is represented by a purple hollow cylinder in the figure, and the thickness of the overburden between the top of the lining and the surface is $17m$. A 1:1 size detector is placed in the model, and the red cone in the figure is the detection range of the detector. It can be seen from the figure that the observation range of the two observation points has a large combination, indicating that the entire line can be scanned at this observation distance interval. In addition, since the subway line is located directly below Xiaoying West Road, there is no influence of pavement buildings above the observation sites, which is conducive to simulation verification.

\begin{figure}[ht]
\centering
    \begin{minipage}[t]{0.5\linewidth}
        \centering
        \includegraphics[width=\textwidth]{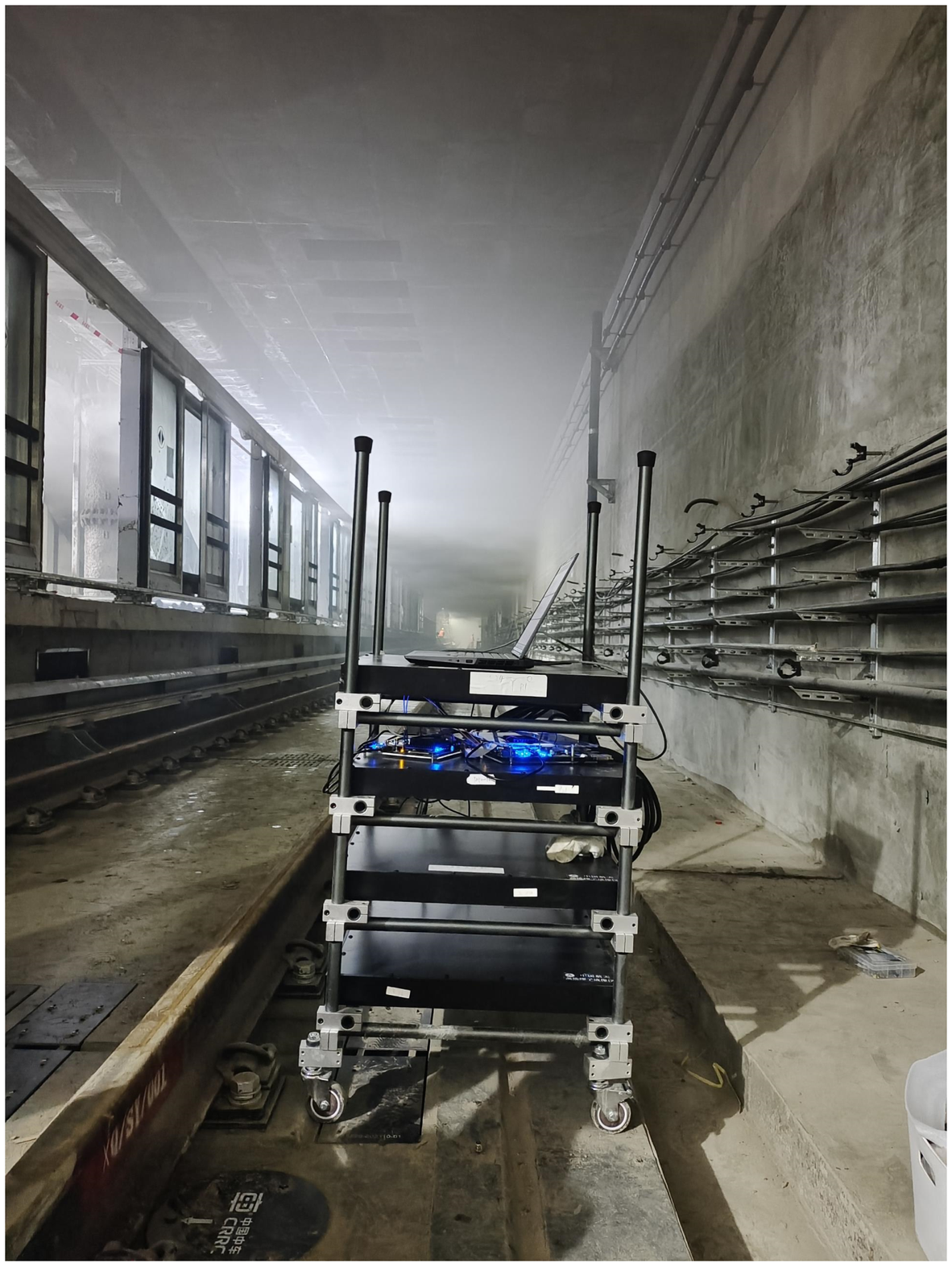}
        \centerline{(a)}
    \end{minipage}%
    \begin{minipage}[t]{0.5\linewidth}
        \centering
        \includegraphics[width=\textwidth]{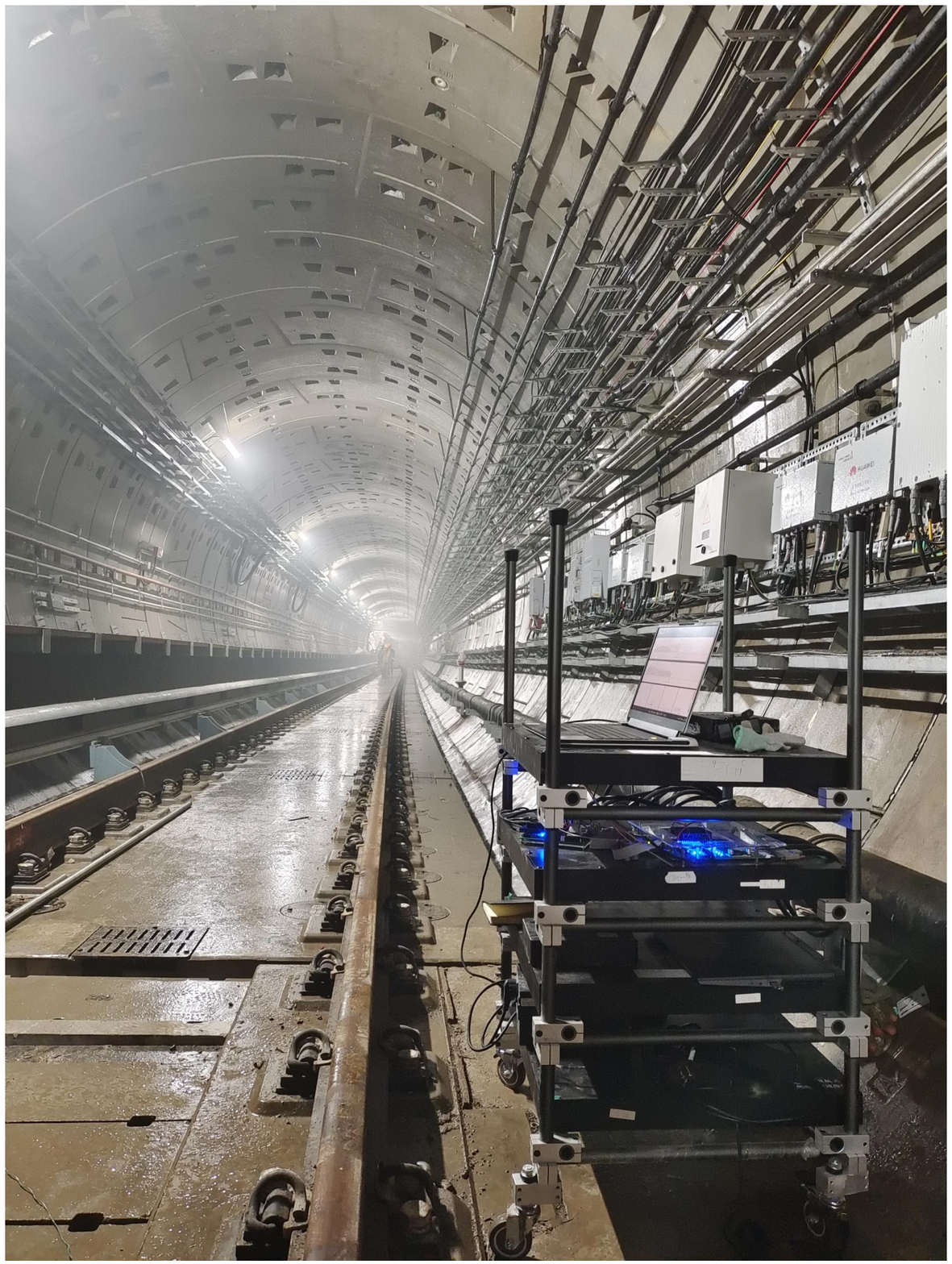}
        \centerline{(b)}
    \end{minipage}
    \caption{\label{fig:scene}Detection scenes in the platform section (a)  and tunnel section (b).}
\end{figure}

In order to eliminate the error of the detector itself and measure the stability of the detector, it is necessary to conduct a ground/open-sky experiment at first, as shown by equation \eqref{e2}. Therefore, the detector was placed skyward on the ground and measured for 114 hours before entering the tunnel. After this, the underground observation started. The detector panel is placed horizontally, facing the upper overlay. In order not to affect the engineering vehicles moving on the track, the detector was placed on the north side of the track against the wall, and the detector center is about $0.6m$ from the track. Started at the position of the marble column No. 11 in the middle of the platform (the first red cone point in Figure \ref{fig:CAD}), the detector was moved from west to east for point-by-point scanning, with each point observed for 1 hour. A total of 21 points were observed, the cumulative time was 210 hours and the scanning distance was $264m$. Among them, each point is observed every $9.56m$ in the platform section, a total of 11 points are observed, and the scanning distance is $90.72m$. It is worth noting that observation point 11 is located directly below the interface between the eastern edge of the hall and the tunnel section. A total of 10 points were observed in the tunnel section, with the first 7 points spaced $12m$ apart and the last 3 points spaced $30m$ apart. Figure \ref{fig:scene} shows the observation scenes of the detector in the platform section and tunnel section.

\section{Results}
\subsection{Open-sky muon flux measurement}
\begin{figure}[htbp]
\centering
\includegraphics[width=0.5\textwidth]{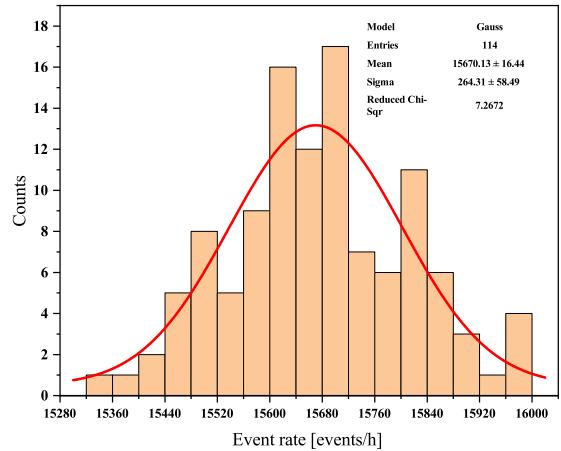}
\caption{\label{fig:opensky}Open-sky histogram distribution and statistical fitting result of the muon rates by 114-hour measurement.}
\end{figure}

The open-sky flux measurement result before entering the subway tunnel is shown in Figure \ref{fig:opensky}. The histogram shows the statistical distribution and the Gaussian fitting result of muon rates obtained from 114-hour measurement. It can be seen that the observed muon rate under the open-sky case is $15670\pm 264 h^{-1}$, and the data stability of the detector is $1.69\%$.
% The left figure shows the position distribution of the muon events according to the hourly normalization. It can be seen that there are more events near vertical at the center position, while there are relatively few events with a large angle inclination at the edges. The right figure shows the histogram of the distribution of muon rates and the Gaussian fitting results obtained from 114-hour measurement results. It can be seen that the observed muon rate under the open-sky case is \textcolor{red}{$15670\pm 264 h^{-1}$}, and the data stability of the detector is $1.69\%$.

% \begin{figure*}[htbp]
% \centering
%     \begin{minipage}[t]{0.38\linewidth}
%         \centering
%         \includegraphics[width=\textwidth,height=6cm]{opensky_dis.eps}
%         \centerline{(a)}
%     \end{minipage}%
%     \begin{minipage}[t]{0.5\linewidth}
%         \centering
%         \includegraphics[width=\textwidth,height=6cm]{opensky_his_en1.eps}
%         \centerline{(b)}
%     \end{minipage}
%     \caption{\label{fig:opensky}Left: Position distribution of open-sky muon events according to the hourly normalization. Right: Histogram distribution and statistical fitting results of 114-hour measurement results.}
% \end{figure*}

\subsection{Underground muon flux measurement}
\begin{figure*}[htbp]
\centering
\includegraphics[width=0.8\textwidth, height = 5cm]{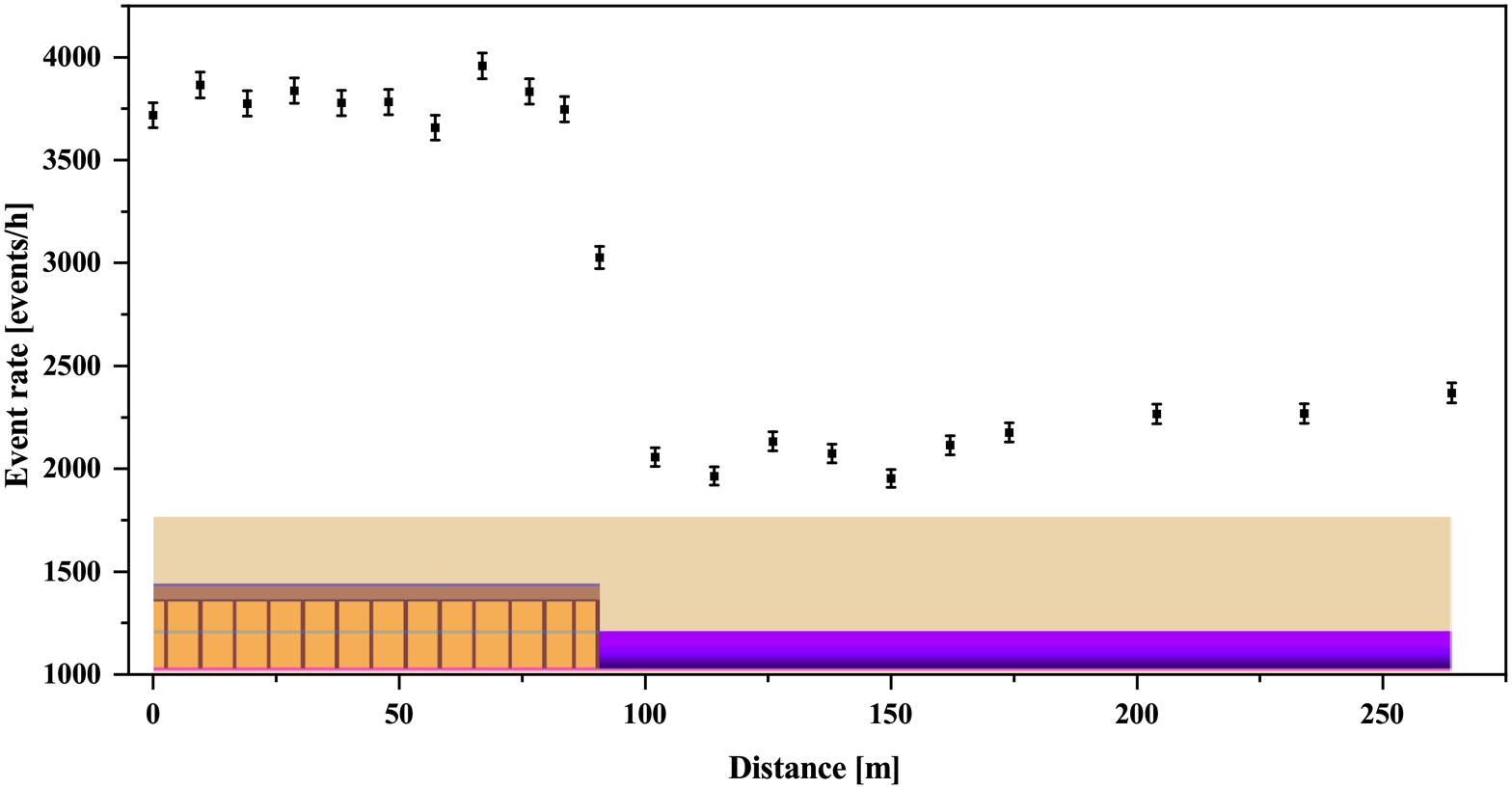}
\caption{\label{fig:flux_dis_underground}Underground measurement results of muon rates and corresponding statistical errors as a function of distance.}
\end{figure*}

Muon event rate measurements at 21 locations within the platform section and tunnel section are shown in Figure \ref{fig:flux_dis_underground}, of which the error bars represent the statistical errors and the subway tunnel model profile is drawn at the bottom of the diagram according to the coordinate relationship. It can be seen that the average muon rates are $3795.4 \pm 62.68h^{-1}$ in the platform section, $3027\pm 55 h^{-1}$ at the intersection face and $2137.3 \pm 105.96 h^{-1}$ in the tunnel section.
% Furthermore, since the distributions of muon flux at different locations in each section are basically the same, the results of two representative locations are selected for simplicity, with the flux distribution of location 3 representing those of the remaining locations in the platform section and the flux distribution of location 15 representing those of the remaining locations in the tunnel section, as shown in Figure \ref{fig:position_dis_underground}. The detailed observation results of the remaining locations are shown in the appendix. 
% \vspace{-0.8cm} %调整图片与上文的垂直距离
\begin{figure*}[htbp]
\centering
\includegraphics[width=0.8\textwidth, height =14cm ]{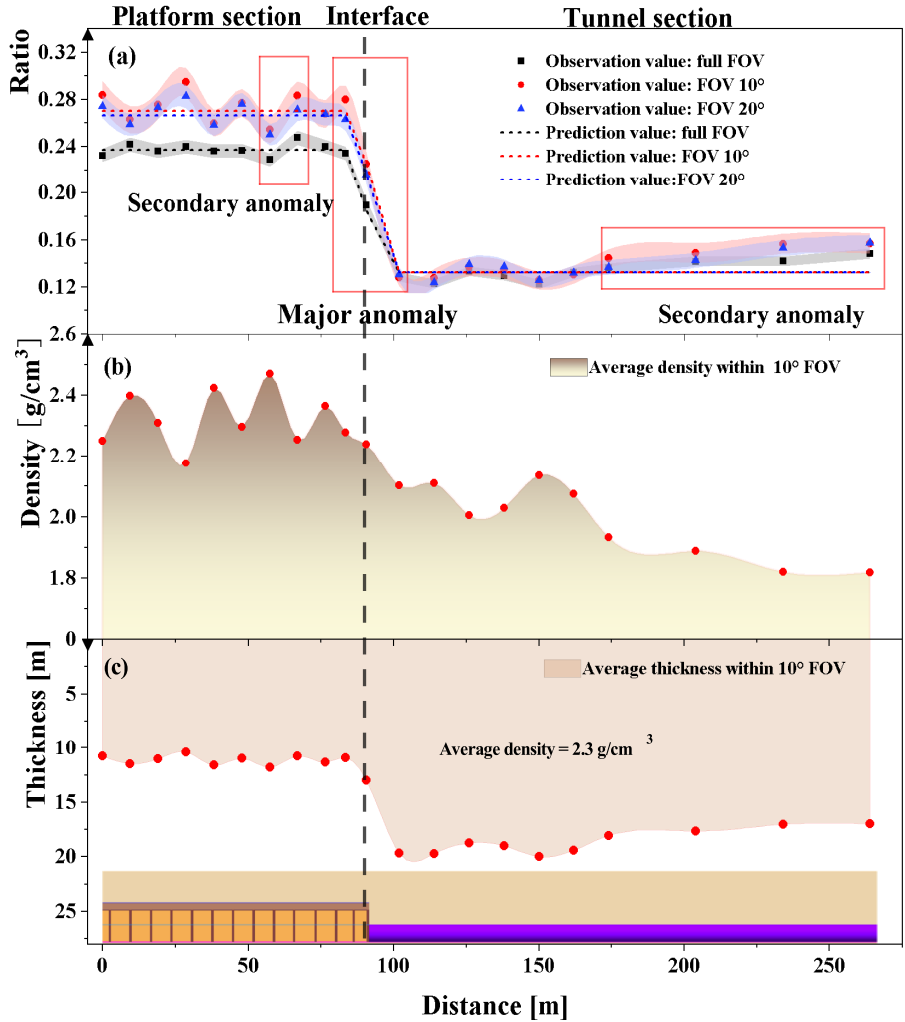}
\caption{\label{fig:position_dis}The observed and predicted ratios vary with distance (a); Assuming that the structure size is fixed, the average density of the 10-degree-cone overburden varies with distance (b).Assuming an average density of $2.3 g/cm^3$, the average thickness of the 10-degree-cone overburden varies with distance (c).}
% (Black square and black area: ratio and error band for all observation angle ranges; Red circle and red area: ratio and error band within the range of nearly 10 degrees vertical cone angle; Blue triangle and blue area: rate and error band within the range of nearly 20 degrees cone angle
\end{figure*}

% \vspace{-0.8cm} %调整图片与上文的垂直距离

% \begin{figure*}[htbp]
% \centering
%     \begin{minipage}[t]{0.3\linewidth}
%         \centering
%         \includegraphics[width=\textwidth]{point3.eps}
%         \centerline{(a)}
%     \end{minipage}%
%     \begin{minipage}[t]{0.3\linewidth}
%         \centering
%         \includegraphics[width=\textwidth]{point11.eps}
%         \centerline{(b)}
%     \end{minipage}
%     \begin{minipage}[t]{0.3\linewidth}
%         \centering
%         \includegraphics[width=\textwidth]{point15.eps}
%         \centerline{(c)}
%     \end{minipage}
%     \caption{\label{fig:position_dis_underground}Underground measurements for the position distribution of muon events at (a) location 3, (b) location 11, (c) location 15.}
% \end{figure*}

% \begin{figure*}[htbp]
% \centering
% \includegraphics[width=0.8\textwidth]{position_dis_en.eps}
% \caption{\label{fig:underground}Underground measurements for the position distribution of muon rate (up), the polar coordinate distribution of ratio (middle), and the polar coordinate distribution of ratio by 3×3 sliding window (down) for the platform section, interface and tunnel section.}
% \end{figure*}

\subsection{Ratio, density and thickness}
The observed and predicted values of the ratios at 21 locations within the platform section and tunnel section are distributed as a function of distance, as shown in Figure \ref{fig:position_dis}(a). The data points in different colors are measured ratios within different observation angle ranges, and the corresponding color ramp range is the statistical error of the data. The dotted lines corresponding to the colors in the figure are the predicted ratios obtained from the Geant4 simulation using the CAD model in Figure \ref{fig:CAD}. It can be seen from the figure that there is a big difference in the ratio points within the platform section and the tunnel section in the figure, point 11 is located at the intersection interface and its ratio is in the middle position. In addition, the ratio value of the tunnel section has a tendency to increase. Secondly, by comparing with the predicted values, points 7 and 8 deviate greatly from the simulated data, indicating that there are anomalies in the corresponding observation range, resulting in a low observed number at location 7 and a large number at location 8.

% \subsection{Average density varies with distance SH}
According to the relationship between ratio and density length in Figure \ref{fig:ratio-rho}, since the structural dimensions of the subway station and tunnel are known, assuming that the thickness of the platform and tunnel part of the overburden remains constant, using the observation results in the near vertical direction of 10 degrees, the average density distribution of the overburden above the detection location is inverted as a function of distance, as shown in Figure \ref{fig:position_dis}(b).

% \subsection{Average thickness varies with distance}
Similarly, assuming that the average density for the overburden above the detector is fixed at $2.3g/cm^3$, the average value of the thickness of the overburden can be calculated according to the density length. The average thickness distribution of the 10 degree cone angle overburden above the detector is obtained as shown in Figure \ref{fig:position_dis}(c). It can be seen that the thickness of overburden within the platform section is relatively stable, with an average thickness of $11.11m$. The average thickness of the overburden within the tunnel section is $18.5m$, and there is a trend of gradual thinning.

\section{Discussion}
\subsection{Anomaly Analysis}
As can be seen from Figure \ref{fig:position_dis}(a), the change in the muon ratio detected by the detector is highly sensitive to anomalies. Due to the characteristics of large geometric acceptance of detector in the vertical direction but small in the inclined direction, the muons with large inclination angle would have fewer event statistics and large statistical errors. In order to reduce the influence of the muon with a large angle tilt, the muon events in the range of 10 degrees, 20 degrees, and all observation angles were selected and analyzed separately, with muons in the 10-degree cone angle range was approximately considered to be vertical. According to the observation results in Figure \ref{fig:position_dis}(a), the major anomaly is the difference between the platform section and the tunnel, where the ratio values at observation locations in the two sections in the figure are quite different. Secondary anomalies within the platform section and tunnel section can be analyzed separately. In the platform section, under the full observation range of the detector, the observed values of points 7 and 8 deviate from the predicted values and are outside the error range. Within the tunnel section, the observed values deviate from the predicted values and the amount of deviation gradually increase with distance. Here, we mainly discuss these three part of anomalies, other abnormalities caused by tiny structures in the subway station overburden or other reasons will not be discussed. Such as point 4, the ratio observation value within the full observation angle range is consistent with the predicted value. However, when the angle range decreases, there is an anomaly. It is guessed that there is a density loss caused by structure such as cable channels directly above this location. Due to its small size, it will not be discussed in the article.

For the major anomaly, the difference in the ratios between the platform section and the tunnel is very significant, nearly doubling. The interface ratio value is somewhere in between. Since the experimental observations and simulated predictions in the Figure \ref{fig:position_dis}(a) are basically consistent within the statistical error range, it can be seen from the structure of the subway model that the anomaly is caused by the density loss of the lobby, around $8m$ in height, on the first floor of the station, which is known from the architectural drawings.
% In addition, it can be seen from Figure \ref{fig:underground} that there are obvious differences in the position distribution of the muon events in the platform section, the intersection interface, and the tunnel section. In the platform section, the detector is placed against the wall on the north side, while on the south side it's empty. Affected by the wall, the muon rate and ratio distribution are obviously asymmetrical, with more events on the south side and fewer events on the north side. At the interface, the west side is the platform section hall, and its total overburden thickness is smaller than the east side. The corresponding muon rate and ratio distribution show left and right asymmetry, with more events on the left and less on the right. The overburden above the tunnel section is relatively uniform, so the muon rate distribution is also symmetrical.

\begin{figure*}[htbp]
\centering
    \begin{minipage}[t]{0.46\linewidth}
        \centering
        \includegraphics[width=\textwidth,height=5cm]{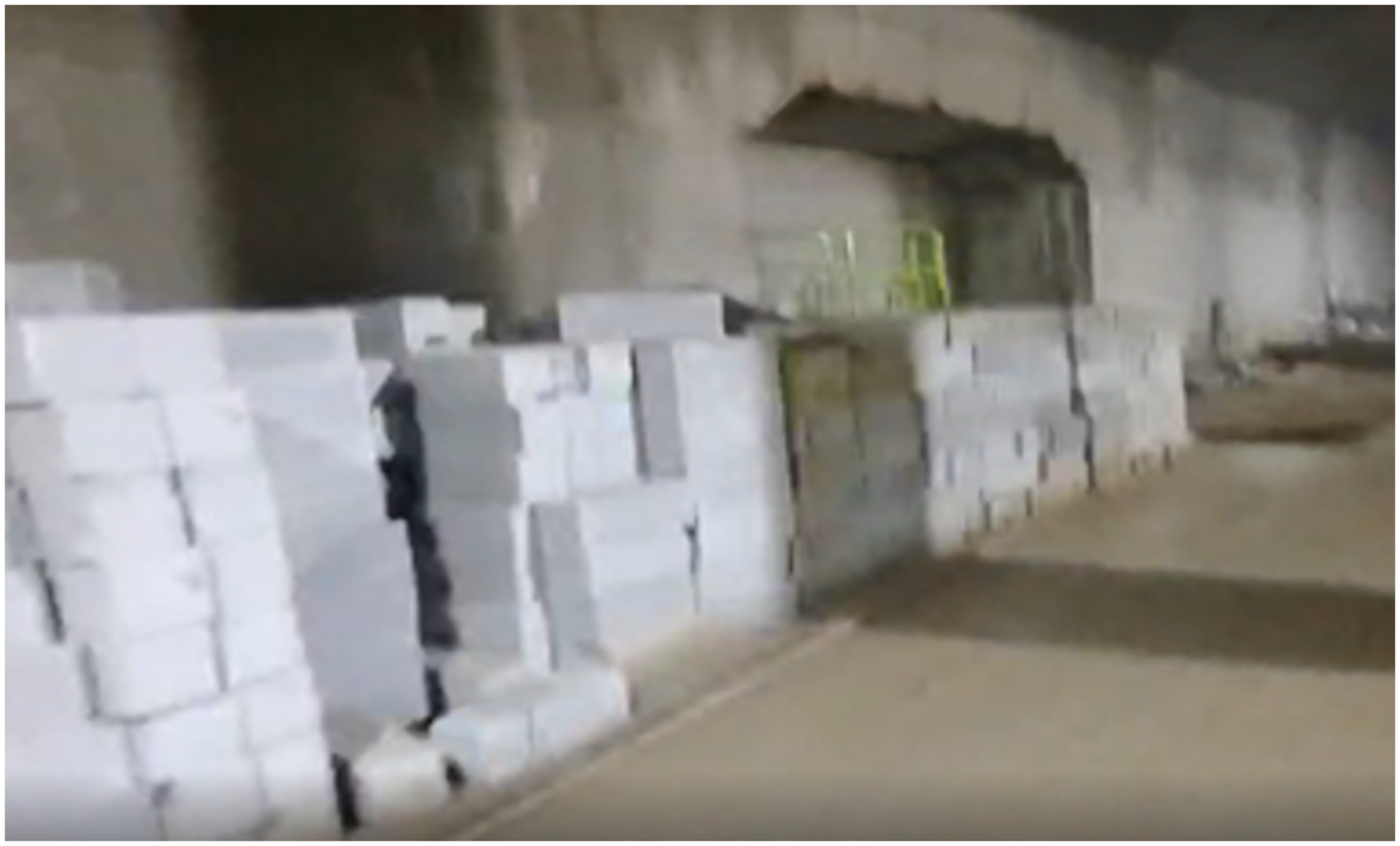}
        \centerline{(a)}
    \end{minipage}%
    \quad
    \begin{minipage}[t]{0.46\linewidth}
        \centering
        \includegraphics[width=\textwidth,height=5cm]{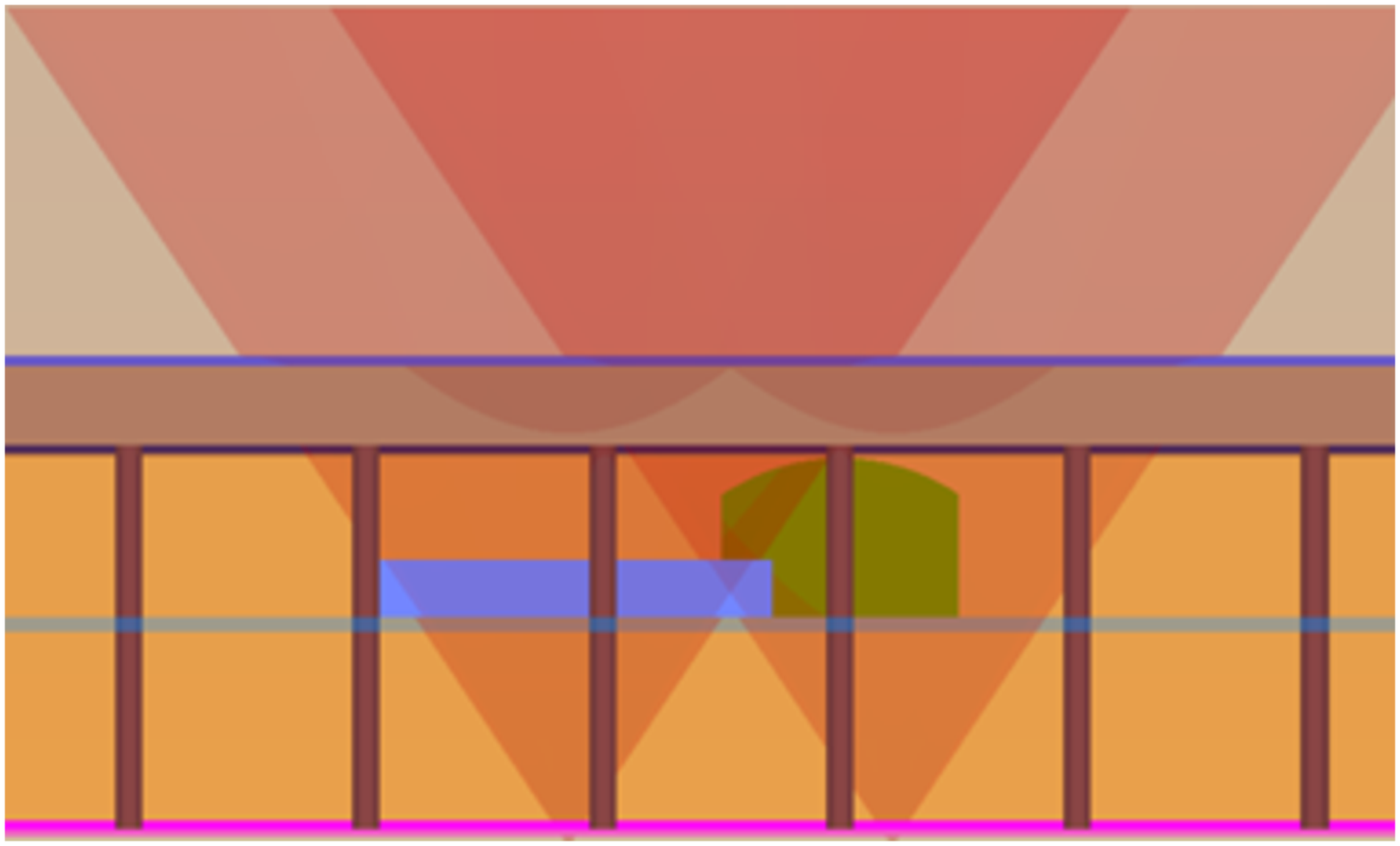}
        \centerline{(b)}
    \end{minipage}
    \caption{\label{fig:anomaly}Light body brick pile and passage mouth above locations 7 and 8 (a). Model diagram of the light body brick pile and passage mouth (b).}
\end{figure*}

\begin{figure}[htbp]
\centering
\includegraphics[width=0.5\textwidth]{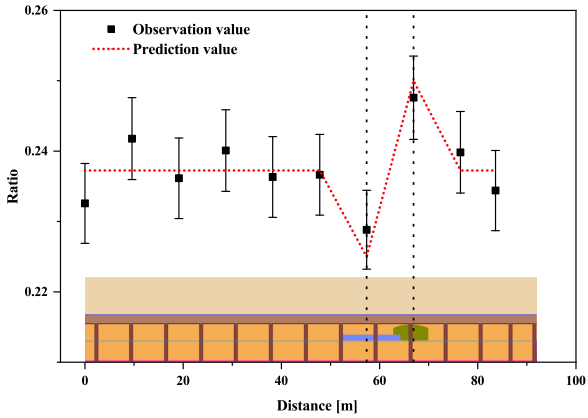}
\caption{\label{fig:anomaly_dis}Simulation results for the anomalies at locations 7 and 8 in the platform section.}
\end{figure}

For the secondary anomalies in the platform section, according to Figure \ref{fig:position_dis}(a), there are two anomalies by the full-range observation, location 7 and 8. As the observation angle range decreases, the anomaly at location 7 has always existed, while the anomaly at location 8 has gradually disappeared, indicating that the anomaly at location 7 may be located directly above the detector, while the anomaly at location 8 is above the side of the detector. In the lobby on the first floor within the platform section, it was observed that there was a pile of light bricks above location 7, and a passage mouth with a depth of about $2m$ on the wall side above above location 8, as shown in Figure \ref{fig:anomaly}(a). Consulting the engineering drawings of the subway station shows that the passage entrance is the entrance of the subway station passage B, which is currently in the stage of unexcavated. The lightweight brick pile makes the muon flux at location 7 small, while the passage mouth makes the muon flux at location 8 large, resulting in abnormalities. In order to confirm the guess of the anomaly at locations 7 and 8, the light brick pile and the passage mouth were added to the original subway tunnel model in Figure \ref{fig:CAD} for simulation, as shown in Figure \ref{fig:anomaly}(b). The light body brick pile was $1.6m$ high, $14m$ long, $2.8m$ wide, and the density was set to $1.7g/cm^3$. The entrance of passage B is $7m$ wide and $3.6m$ high, and is topped by a round arch with a radius of $6.5m$. Through the Geant4 simulation, the prediction value of ratio in the full observation angle range is shown by the dotted line in figure \ref{fig:anomaly_dis}. It can be seen that the predicted values of the anomalies at locations 7 and 8 are consistent with the actual observed values.

For secondary anomalies in the tunnel section, it can be seen from Figure \ref{fig:position_dis}(a) that from the position of $174m$, the observed ratio within the tunnel section gradually becomes larger and deviates from the predicted value. It is speculated that the overburden above the tunnel section is gradually thinning. Due to the lack of actual data on the thickness of the overburden in the tunnel section, further observation of the muons at longer time and denser points are required to confirm this hypothesis.

\subsection{Error source analysis}
For the observation values, the detector itself has good stability and the cosmic ray muon flux fluctuates little \cite{altitude2010} at the same altitude, under the same pressure and temperature conditions. Moreover, for the muon radiography of large scale target, the effect of multiple coulomb scattering is negligible \cite{scatter_2022}. Therefore, the system error of detector is neglected \cite{Li_2022} and only statistical error is considered in the paper. The statistical error of the muon rate is estimated using $\sigma=1/\sqrt{N}$, and the ratio error is calculated using the error transfer formula of division, as shown in the error band of Figure \ref{fig:position_dis}(a). For the inverted average density (Figure \ref{fig:position_dis}(b)) and average thickness (Figure \ref{fig:position_dis}(c)) of the overburden, the main error comes from the inaccuracy of the structural CAD model as well as the assumed density value. Since the subway station in the actual situation is much more complex than the model, there are many structures in the overburden that are relatively small compared to the light brick pile and the passage mouth, such as ventilation ducts, cable sandwiches, etc., which will cause the thickness and density of the inversion to deviate from the actual value more or less. In addition, since the relationship between ratio and density length is used to derive the average density and average thickness (Figure \ref{fig:ratio-rho}), the error caused by this conversion relationship, whether estimated using empirical formulas or simulated by software packages, cannot be ignored. Therefore, the anomaly analysis is mainly carried out by comparing the difference between the observed value and the predicted value of the ratio, and the average thickness and average density distribution of the conversion are only used as reference.

\section{Conclusion}
Muography is an innovative way to study the internal structure of large buildings  non-intrusively, non-destructively and conveniently. Nowadays, muography has many successful research cases in volcanoes, ancient building archaeology, nuclear safety and other aspects at both home and abroad. The main principle of muography is to use the muon detector to observe the change of muon rate before and after muons penetrate the object, and reflect the density or thickness inside the target according to the relationship between the survival ratio of the muon rate and density length of the matter. For subway tunnels, structural anomalies such as cavities in the overburden are safety factors that have to be considered. Compared with traditional detection methods, muography has no damage and deep detection thickness, which is more suitable for the detection of the overburden of urban subway tunnels. In this paper, a muon detector composed of four layers of unit plates sized $50cm\times50cm$ based on plastic scintillator and SIPM readout is developed, with a pixel size of $2cm$. The detector has a broad angular aperture of about $79^{\circ}$, a high average angular resolution of $33.3mrad$, and a  maximum acceptance of $2.78cm^{2}sr$. The detector is equipped with a $2000Wh$ mobile power supply and pulleys at the bottom, which is convenient for moving and continuous acquisition in the tunnel. Based on this detector, this paper conducts a muon rate scanning experiment for the overburden of Xiaoying West Road railway station under construction in Beijing, and analyzes the main structural anomalies.

The first measurements were made on the ground before the underground experiment. 114 hours of data have been obtained in the case of open sky and the data stability was $1.69\%$. The underground experiment can be divided into two parts, the platform section where there is a two-story hall inside the station and the tunnel section outside the station. The detector scanned the overburden along the track eastward from the mid-position of the platform section, observing at each location for 1 hour. A total of 21 points were scanned, with a total length of about $264m$.

By comparing the observed and predicted values of the ratios, the detected anomalies are divided into major anomaly, secondary anomalies in the platform section, and secondary anomalies in the tunnel section. The major anomaly was caused by the lobby on the first floor of the subway station in the platform section, resulting in the ratio values within the platform section being nearly double those within the tunnel section. As for the secondary anomalies in the platform section, the main ones are at locations 7 and 8 where the ratios deviate from the predicted values within the full observation angle range. The ratio value at location 7 is low and does not change with the narrowing of the observation angle range. The ratio value at location 8 is large, and the difference gradually decreases as the observation angle range decreases. Through observation, the speculative anomalies were found in the lobby on the first floor, which were the pile of light bricks directly above location 7 and the passage mouth above the side of location 8. In order to confirm this conjecture, two structures, the light brick pile and the passage mouth, were added to the tunnel model according to their actual sizes. The Geant4 simulation results were consistent with the observed values within the error range. For the tunnel section, the ratio observation deviates from the predicted value from the location at $174m$ and gradually becomes larger, it is guessed that the overburden of the tunnel in this section gradually becomes thinner. If the anomaly is to be further analyzed, further long-term, denser-point observation is required.

The systematic error is independent of the statistic and is relatively small and negligible. For the muon rate and ratio observations, only the impact of the statistical error is considered, and the estimation is made using $\sigma=1/\sqrt{N}$. For the inverted average density and average thickness of the overburden, the main error comes from the inaccuracy of the model size and the assumed density. Secondly, the error from the conversion relationship between ratio and density length cannot be ignored either.

In general, through this muon observation experiment, not only the density loss caused by the lobby located on the first floor of the platform section was accurately detected, but also secondary anomalies such as light brick piles and unexcavated passage mouth were detected. It shows that the muon event rate is very sensitive to the change of the structure of the overburden layer, and muography may be a very promising technology for tunnel safety detection.

\section{Acknowledge}
This work is supported by National Natural Science Foundation of China (Grant No. 42174076, 41974064 and U1865206), Xi'an Research Institute of China Coal Technology \& Engineering Foundation (Grant No. 2021XAYJC01). We also express our gratitude to YanHeng Li for his great recommendations of experiment site.

% \bibliographystyle{unsrt}
% \bibliography{sample}

\end{document}